\documentclass[aps,prl,preprint,tightenlines,superscriptaddress,showpacs,byrevtex]{revtex4}
\usepackage{graphicx} 
\usepackage{dcolumn}  

\graphicspath{{ps}}

\def\Journal#1#2#3#4{{#1} {\bf #2}, #3 (#4)}

\def\NPB{{\em Nucl. Phys.} B}
\def\PLB{{\em Phys. Lett.}  B}
\def\PRL{\em Phys. Rev. Lett.}
\def\PRD{{\em Phys. Rev.} D}
\def\ZPC{{\em Z. Phys.} C}

\def\RMP{\em Rev. Mod. Phys.}

\def\Kstll{K^* \ell^+ \ell^-}
\def\Kll{K \ell^+ \ell^-}
\def\Xsll{X_s \ell^+ \ell^-}
\def\Xs2ll{X_s \ell \ell}
\def\KKstll{K^{(*)} \ell^+ \ell^-}

\def\Ksthh{K^* h^+ h^-}
\def\Kstarhh{K^* h h}
\def\Kstlh{K^* \ell^{\pm} h^{\mp}}
\def\Kstarlh{K^* \ell h}

\def\Mbc{M_{\rm bc}}

\def\sig{{\rm sig}}

\def\ol{\overline}
\def\PM#1#2{\,^{+#1}_{-#2}}

\begin{document}


\preprint{\vbox{ \hbox{   }
                 \hbox{BELLE-CONF-0521}
                 \hbox{LP2005-156}
                 \hbox{EPS05-493} 
}}

\title{ \quad\\[0.5cm]  Measurement of Forward-Backward Asymmetry and Wilson Coefficients in $B \to \Kstll$}

\affiliation{Aomori University, Aomori}
\affiliation{Budker Institute of Nuclear Physics, Novosibirsk}
\affiliation{Chiba University, Chiba}
\affiliation{Chonnam National University, Kwangju}
\affiliation{University of Cincinnati, Cincinnati, Ohio 45221}
\affiliation{University of Frankfurt, Frankfurt}
\affiliation{Gyeongsang National University, Chinju}
\affiliation{University of Hawaii, Honolulu, Hawaii 96822}
\affiliation{High Energy Accelerator Research Organization (KEK), Tsukuba}
\affiliation{Hiroshima Institute of Technology, Hiroshima}
\affiliation{Institute of High Energy Physics, Chinese Academy of Sciences, Beijing}
\affiliation{Institute of High Energy Physics, Vienna}
\affiliation{Institute for Theoretical and Experimental Physics, Moscow}
\affiliation{J. Stefan Institute, Ljubljana}
\affiliation{Kanagawa University, Yokohama}
\affiliation{Korea University, Seoul}
\affiliation{Kyoto University, Kyoto}
\affiliation{Kyungpook National University, Taegu}
\affiliation{Swiss Federal Institute of Technology of Lausanne, EPFL, Lausanne}
\affiliation{University of Ljubljana, Ljubljana}
\affiliation{University of Maribor, Maribor}
\affiliation{University of Melbourne, Victoria}
\affiliation{Nagoya University, Nagoya}
\affiliation{Nara Women's University, Nara}
\affiliation{National Central University, Chung-li}
\affiliation{National Kaohsiung Normal University, Kaohsiung}
\affiliation{National United University, Miao Li}
\affiliation{Department of Physics, National Taiwan University, Taipei}
\affiliation{H. Niewodniczanski Institute of Nuclear Physics, Krakow}
\affiliation{Nippon Dental University, Niigata}
\affiliation{Niigata University, Niigata}
\affiliation{Nova Gorica Polytechnic, Nova Gorica}
\affiliation{Osaka City University, Osaka}
\affiliation{Osaka University, Osaka}
\affiliation{Panjab University, Chandigarh}
\affiliation{Peking University, Beijing}
\affiliation{Princeton University, Princeton, New Jersey 08544}
\affiliation{RIKEN BNL Research Center, Upton, New York 11973}
\affiliation{Saga University, Saga}
\affiliation{University of Science and Technology of China, Hefei}
\affiliation{Seoul National University, Seoul}
\affiliation{Shinshu University, Nagano}
\affiliation{Sungkyunkwan University, Suwon}
\affiliation{University of Sydney, Sydney NSW}
\affiliation{Tata Institute of Fundamental Research, Bombay}
\affiliation{Toho University, Funabashi}
\affiliation{Tohoku Gakuin University, Tagajo}
\affiliation{Tohoku University, Sendai}
\affiliation{Department of Physics, University of Tokyo, Tokyo}
\affiliation{Tokyo Institute of Technology, Tokyo}
\affiliation{Tokyo Metropolitan University, Tokyo}
\affiliation{Tokyo University of Agriculture and Technology, Tokyo}
\affiliation{Toyama National College of Maritime Technology, Toyama}
\affiliation{University of Tsukuba, Tsukuba}
\affiliation{Utkal University, Bhubaneswer}
\affiliation{Virginia Polytechnic Institute and State University, Blacksburg, Virginia 24061}
\affiliation{Yonsei University, Seoul}
  \author{K.~Abe}\affiliation{High Energy Accelerator Research Organization (KEK), Tsukuba} 
  \author{K.~Abe}\affiliation{Tohoku Gakuin University, Tagajo} 
  \author{I.~Adachi}\affiliation{High Energy Accelerator Research Organization (KEK), Tsukuba} 
  \author{H.~Aihara}\affiliation{Department of Physics, University of Tokyo, Tokyo} 
  \author{K.~Aoki}\affiliation{Nagoya University, Nagoya} 
  \author{K.~Arinstein}\affiliation{Budker Institute of Nuclear Physics, Novosibirsk} 
  \author{Y.~Asano}\affiliation{University of Tsukuba, Tsukuba} 
  \author{T.~Aso}\affiliation{Toyama National College of Maritime Technology, Toyama} 
  \author{V.~Aulchenko}\affiliation{Budker Institute of Nuclear Physics, Novosibirsk} 
  \author{T.~Aushev}\affiliation{Institute for Theoretical and Experimental Physics, Moscow} 
% ...
\collaboration{The Belle Collaboration}

\noaffiliation

\begin{abstract}
We report a measurement of the ratios of Wilson coefficients $A_9/A_7$ and $A_{10}/A_7$ in
$B \to \Kstll$. The result is obtained from a  data sample containing 386
million $B\bar{B}$ pairs that was collected 
at the $\Upsilon(4S)$ resonance
with the Belle detector at the KEKB asymmetric energy $e^+ e^-$
collider.
\end{abstract}

\pacs{13.20.He, 11.30.Hv}

\maketitle

\tighten

{\renewcommand{\thefootnote}{\fnsymbol{footnote}}}
\setcounter{footnote}{0}

Flavor-changing neutral current (FCNC) $b \to s$ processes are forbidden at
tree level in the Standard Model (SM); rather, they proceed at a low
rate via loop or box diagrams. If additional diagrams with
non-SM particles contribute to such a decay, their amplitudes will
interfere with the SM amplitudes and thereby modify the decay rate as
well as other properties.
This feature makes FCNC processes an ideal place to search for new physics. 

To evaluate the new physics contributions in $b \to s$ processes,
Wilson coefficients are used~\cite{BBL}. These coefficients
parameterize the strength of the short distance interaction. If new physics
contributes to the $b \to s$ processes, the relevant coefficients will deviate
from the SM values. For electroweak penguin decays, the effective Wilson coefficients 
$C_7^{\rm eff}$, $C_9^{\rm eff}$ and $C_{10}^{\rm eff}$ appear in the partial decay width.
A next-to-next-to-leading order~(NNLO) calculation for these effective coefficients has
many correction terms~\cite{WC}, so leading coefficients $A_7$, $A_9$ and $A_{10}$ are usually used
for the evaluation.

Measurements of the radiative penguin decay $B\to
X_{s}\gamma$~\cite{belle-btosgamma,cleo-btosgamma,aleph-btosgamma},
which are consistent with the SM prediction, strongly constrain the
magnitude---but not the sign---of the Wilson coefficient $A_7$.
This is an important constraint; however, non-SM contributions can change
the sign of $A_7$ without changing the $B\to
X_{s} \gamma$ branching fraction~\cite{C7}.

\begin{figure}[htbp]
\begin{center}
  \includegraphics[scale=0.8]{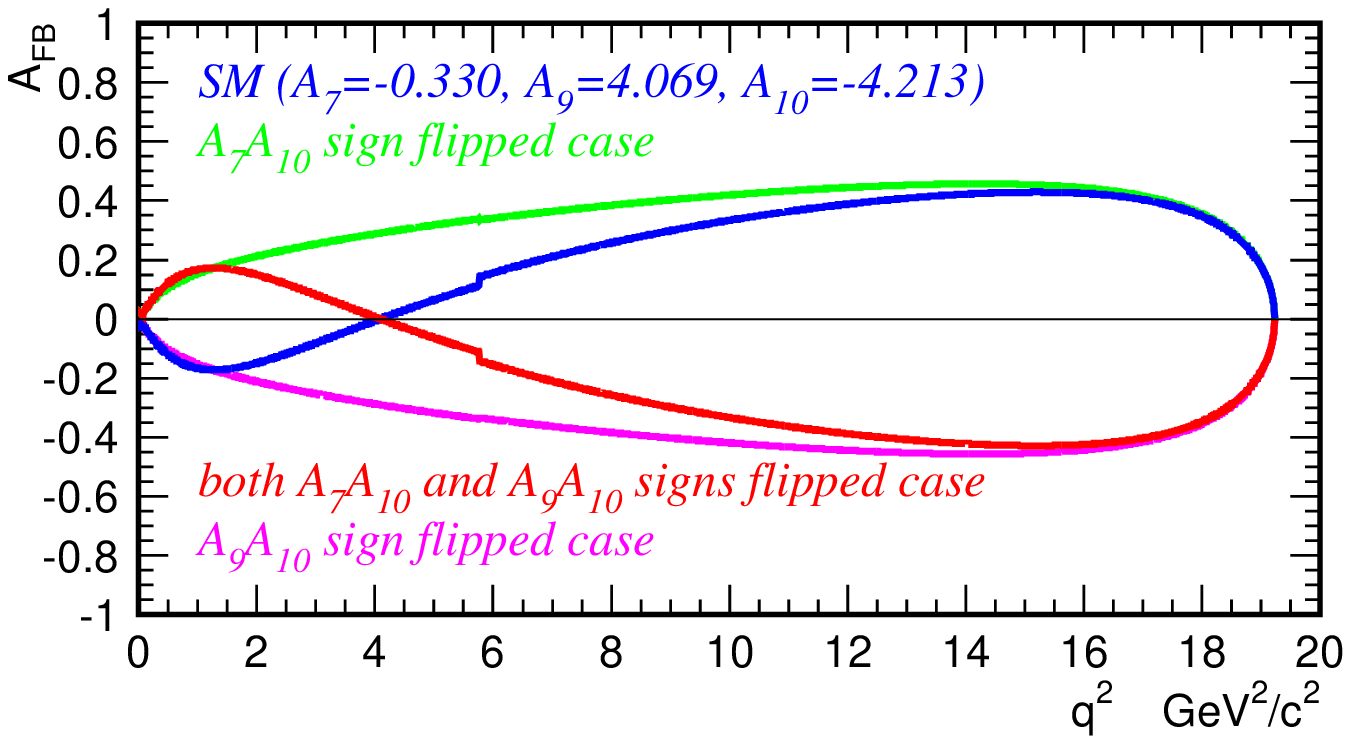}
  \includegraphics[scale=0.8]{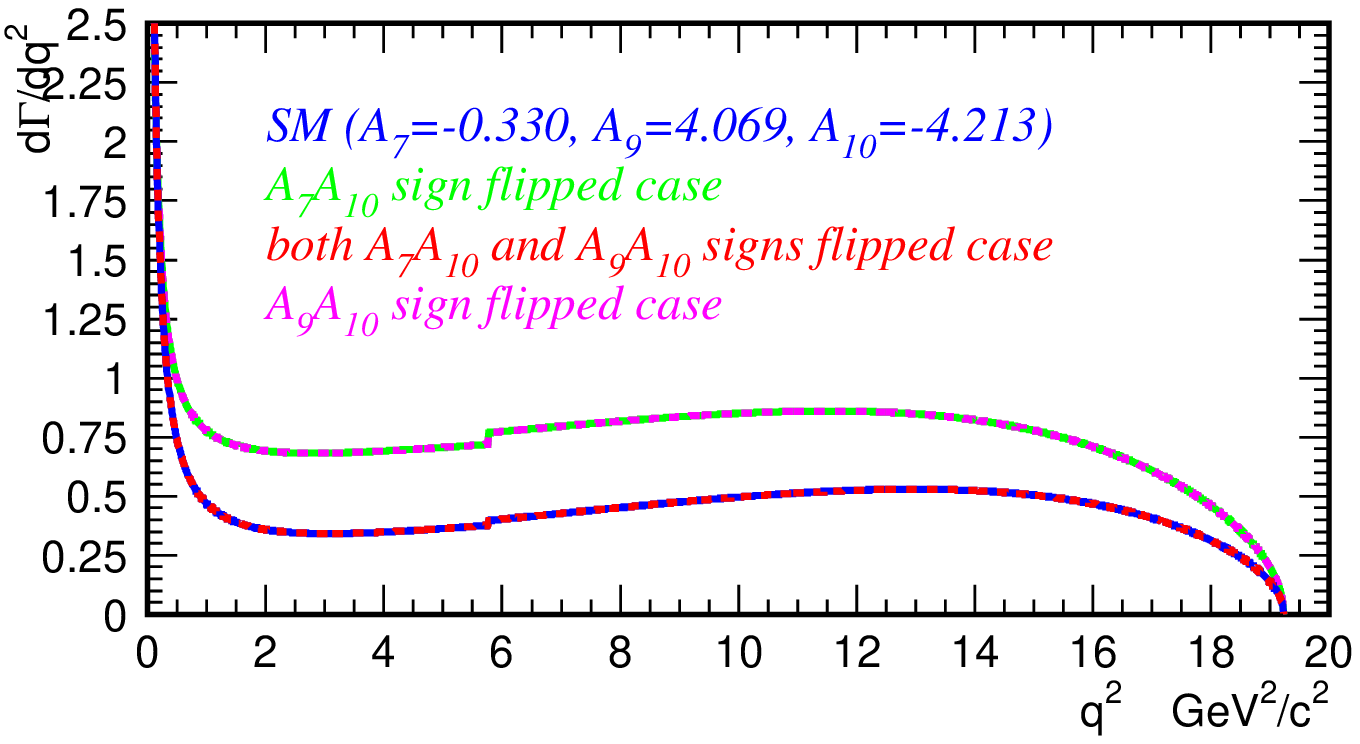}
\end{center}
\caption{Expected forward-backward asymmetry and $q^2$ distributions for several input parameters. The discontinuity at  5.5~GeV/$c^2$ is due to the imperfect matching between full and partial NNLO calculations.}
\label{fig:afbintro}
\end{figure}

The $b \to s \ell^{+} \ell^{-}$ process is promising from this point
of view, since not only the photonic penguin diagram but also the
$Z$-penguin and box diagrams contribute to this decay mode. As a
result, we can determine the relative signs of the Wilson coefficients $A_7$, $A_9$ and $A_{10}$ as well
as their absolute values. The first observations of 
$B\to K \ell^+\ell^-$, $B\to K^* \ell^+\ell^-$ and inclusive $B \to X_{s} \ell^+
\ell^-$ decays were reported by the Belle
Collaboration~\cite{BelleKll,BelleKstarll,BelleXsll}.
The measured branching fractions of these decay modes~\cite{BelleKll,BelleKstarll,BelleXsll,BabarXsll,BelleXsll2} were used to
exclude a large area of the allowed region in the $A_9$-$A_{10}$ plane~\cite{ALGH,Lunghi,Gambino}.
However, the determination of the relative sign of $A_7$ to $A_9$ or $A_{10}$
(as well as $A_9$ and $A_{10}$)
requires precise measurements of differential branching fraction and
the forward-backward asymmetry as functions of $q^2$
in these decay modes~\cite{C7C9C10}. As shown in Fig.~\ref{fig:afbintro}, the distributions are quite sensitive to the choice of Wilson coefficients.

The first measurement of the forward-backward asymmetry in $B \to \KKstll$ was reported by the Belle Collaboration~\cite{BelleAFB}. The result for $B \to \Kstll$ is consistent with both the SM case and the case of a flipped sign for $A_7A_{10}$, however a quantitative evaluation in terms of Wilson coefficients has not been done yet.

In this paper, we present preliminary results of a measurement of 
Wilson coefficients in $B \to \Kstll$ using data produced in $e^+{}e^{-}$
annihilation at the KEKB asymmetric collider \cite{KEKB} and
collected with the Belle detector. 
The data sample corresponds to
357~fb${}^{-1}$ taken at the $\Upsilon(4S)$ resonance and contains
approximately 386 million $B\overline{B}$ pairs. 
As a control sample, we also study the $B^+ \to K^+ \ell^+ \ell^-$ mode which has
very small forward-backward asymmetry regardless of the existence of new physics~\cite{NoAFBKll}.  
Here and throughout the paper,
the inclusion of charge conjugate modes is implied.

%
%

The Belle detector is a large-solid-angle magnetic
spectrometer that
consists of a silicon vertex detector (SVD),
a 50-layer central drift chamber (CDC), an array of
aerogel threshold \v{C}erenkov counters (ACC), 
a barrel-like arrangement of time-of-flight
scintillation counters (TOF), and an electromagnetic calorimeter
comprised of CsI(Tl) crystals (ECL) located inside 
a super-conducting solenoid coil that provides a 1.5~T
magnetic field.  An iron flux-return located outside of
the coil is instrumented to detect $K_L^0$ mesons and to identify
muons (KLM).  The detector
is described in detail elsewhere~\cite{Belle}.
Two inner detector configurations were used. A 2.0 cm beampipe
and a 3-layer silicon vertex detector was used for the first 
140~fb${}^{-1}$ data sample (DS-I) containing 152 million $B\bar{B}$ pairs,
while a 1.5 cm beampipe, a 4-layer silicon detector and a small-cell inner drift chamber were used to record
the remaining 217~fb${}^{-1}$ data sample (DS-II) containing 234 million $B\bar{B}$ pairs~\cite{Ushiroda}.  

%
%

In this analysis, primary charged tracks---except for the daughters
from $K^0_S \to \pi^+ \pi^-$ decays---are required to have impact
parameters relative to the interaction point of less than 5.0~cm along
the $z$ axis (aligned opposite the positron beam) and less than 0.5~cm in the $r\phi$ plane that is transverse to this axis.  This requirement reduces the combinatorial background from photon conversion, beam-gas and beam-wall events. 

Charged tracks are identified as either kaons or pions by a likelihood ratio
based on the CDC specific ionization, time-of-flight information and the
light yield in the ACC.  This classification is superseded for a track that is
identified as an electron or for a pion-like track that is identified as a muon.
Electrons are identified from the ratio of shower energy of the matching ECL cluster to
the momentum measured by the CDC, the transverse shower shape of this cluster, the
specific ionization in the CDC and the ACC response. We require
that the electron momentum be greater than $0.4$~GeV/$c$ to reach the ECL.   
Muons are identified by their penetration depth and transverse scattering
in the KLM. The muon momentum is required to exceed 0.7~GeV/$c$.  The muon
identification criteria are more stringent for momenta below 1.0~GeV/$c$ to suppress
misidentified hadrons. We also apply a kaon veto for electron and muon candidates.

Photons are selected from isolated neutral clusters in the ECL 
with energy greater than $50$~MeV and a shape 
that is consistent with an electromagnetic shower. 
Neutral pion candidates are reconstructed from pairs of
photons, and are required to have an invariant mass 
within $10$~MeV/$c^2$ of the nominal $\pi^{0}$ mass and
a laboratory momentum greater than 0.1 GeV/$c$.
The pion momentum is recalculated by constraining the invariant mass
to the nominal $\pi^0$ mass.
$K^{0}_{S}$ candidates are reconstructed from oppositely 
charged pions that have invariant masses within $15$~MeV/$c^{2}$ 
of the nominal $K^{0}_{S}$ mass. We impose additional criteria based on
the radial impact parameters of the pions ($\delta r$), 
the distance of closest approach of the pions 
along the beam direction ($\delta z$),
the distance of the vertex from the interaction point ($l$), 
and the azimuthal angle difference between the
vertex direction and the $K^{0}_{S}$ momentum direction ($\delta\phi$).
These variables are combined as follows: for $p_{K^{0}_{S}} < 0.5$~GeV/$c$,
we require $\delta z < 8$~mm, $\delta r > 0.5$~mm, and $\delta\phi <0.3$~rad;
for  $0.5$~GeV/$c < p_{K^{0}_{S}} < 1.5$~GeV/$c$, we apply $\delta z < 18$~mm, 
$\delta r > 0.3$~mm, $\delta\phi <0.1$~rad, and $l > 0.8$~mm; 
and for  $p_{K^{0}_{S}} > 1.5$~GeV/$c$, we impose $\delta z < 24$~mm, 
$\delta r > 0.2$~mm, $\delta\phi <0.03$~rad, and $l > 2.2$~mm.

$K^{*}$ candidates are formed by combining 
a kaon and a pion: $K^{+}\pi^{-}$, 
$K^{0}_{S}\pi^{+}$ or $K^{+}\pi^{0}$.
The $K^{*}$ invariant mass is required to lie within 
$75$~MeV/$c^2$ of the nominal $K^{*}$ mass. 
For modes involving neutral pions, 
combinatorial backgrounds
are reduced by the additional requirement
$\cos\theta_{\rm{hel}}<0.8$,
where $\theta_{\rm{hel}}$
is defined as the angle between the opposite of the $B$ momentum
and the kaon momentum direction in the $K^{*}$ rest frame.

$B$ candidates are reconstructed from a $K^{(*)}$ candidate and an
oppositely-charged lepton pair. If a photon with energy less than
500~MeV is found in the 50~mrad cone along the electron or positron momentum
direction, we add the four momentum of the bremsstrahlung photon to the di-electron system.
 We use two variables defined in the CM frame to select $B$
candidates: the beam-energy constrained mass $M_{\rm{bc}} = \sqrt{ {E_{\rm{beam}}^*}^2 -
{p_{B}^*}^{2}}$ and the energy difference
$\Delta E = E_{B}^* - E_{\rm{beam}}^* $, where 
$p_{B}^*$ and $E_B^*$ are the measured momentum and energy, respectively, of the $B$
candidate, and $E_{\rm{beam}}^*$ is the beam energy~\cite{aster}. When multiple candidates are
found in an event, we select the candidate with the smallest value of $|\Delta E|$.

Backgrounds from $B \to J/\psi(\psi') K^{(*)}$ are rejected using the
dilepton invariant mass.
The veto windows are defined as\\
\hspace*{2cm}
\begin{tabular}{r@{}l@{\qquad}l}
  $-0.25~\textrm{GeV/}c^{2}$ & ${}< M_{ee}     - M_{J/\psi}   < 0.07~\textrm{GeV/}c^{2}$ & for $K^{*}$ modes \\
  $-0.20~\textrm{GeV/}c^{2}$ & ${}< M_{ee}     - M_{J/\psi}   < 0.07~\textrm{GeV/}c^{2}$ & for $K$ modes \\
  $-0.20~\textrm{GeV/}c^{2}$ & ${}< M_{ee}     - M_{\psi'} < 0.07~\textrm{GeV/}c^{2}$ & for $K^{*}$ and $K$ modes\\
  $-0.15~\textrm{GeV/}c^{2}$ & ${}< M_{\mu\mu} - M_{J/\psi}   < 0.08~\textrm{GeV/}c^{2}$ & for $K^{*}$ modes\\
  $-0.10~\textrm{GeV/}c^{2}$ & ${}< M_{\mu\mu} - M_{J/\psi}   < 0.08~\textrm{GeV/}c^{2}$ & for $K$ modes\\
  $-0.10~\textrm{GeV/}c^{2}$ & ${}< M_{\mu\mu} - M_{\psi'} < 0.08~\textrm{GeV/}c^{2}$ & for $K^{*}$ and $K$ modes
\end{tabular}\\
If a bremsstrahlung photon is found, we reapply the above vetoes with the invariant mass calculated including this
photon, to reject the background $J/\psi(\psi') \to e^+e^-\gamma(\gamma)$.
For $K^* e^+ e^-$ modes, $B \to J/\psi \, K$ can be a background
if a bremsstrahlung photon is missed and a pion from the other $B$
meson in the event is included.  We suppress this background by the following
prescription: the included pion is discarded and an unobserved bremsstrahlung
photon is added with a direction parallel to the electron or positron and
an energy that gives $\Delta E = 0$ for the $B$ candidate.
If the dilepton mass and the beam-energy constrained mass are consistent
with a $B \to J/\psi \, K$ event, the candidate is vetoed.

We suppress background from photon conversions and $\pi^{0}$ Dalitz
decays by requiring the di-electron mass to satisfy $M_{e^{+}e^{-}} >
0.14~$GeV/$c^{2}$. This cut eliminates possible peaking background from
$B \to K^* \gamma$ and $B \to K^{(*)} \pi^0$.

Background from continuum $q\overline{q}$ events is suppressed using
event topology. Continuum events have a jet-like shape while
$B\overline{B}$ events have a spherical shape in the center-of-mass
frame. 
A Fisher discriminant $\cal{F}$~\cite{fd} is calculated from the energy
flow in 9 cones along the $B$ candidate sphericity axis and the
normalized second Fox-Wolfram moment $R_{2}$~\cite{fw}.  
We combine this with the cosine of the polar angle $\cos\theta_B^*$
of the $B$ meson flight direction 
and the cosine of the polar angle $\cos\theta^{*}_{\rm sph}$ of the $B$ meson sphericity axis to define likelihoods $\cal{L}_{\rm{sig}}$ and
$\cal{L}_{\rm{cont}}$ for signal and continuum background, respectively,
and then cut on the likelihood ratio $\cal{R}_{\rm{cont}} =
\cal{L}_{\rm{sig}} / (\cal{L}_{\rm{sig}} + \cal{L}_{\rm{cont}})$.
For the muon mode, $|\cos\theta^{*}_{\rm{sph}}|$ is not used since its
distribution is nearly the same for signal and continuum within the
detector acceptance.

The dominant background from $B\overline{B}$ events is due to
semileptonic $B$ decays. The missing energy of the event,
$E_{\mathrm{miss}} = 2 E_{\rm beam}^* - E_{\rm vis}^*$ 
where $E_{\rm vis}^*$ is a total visible energy in the event,
is used to suppress this background since 
the undetected neutrinos carry away a substantial amount of energy.
The $B$ meson flight angle, $\cos\theta_B^*$, is also used to suppress combinatorial
background in $B\overline{B}$ events.  
We combine $E_{\mathrm{miss}}$ and $\cos\theta^{*}_{B}$ into signal and
$B\overline{B}$-background likelihoods and cut on the
likelihood ratio ${\cal{R}}_{B\overline{B}}$, defined similarly to
${\cal{R_{\rm{cont}}}}$. 

The signal box is defined as $|M_{\rm bc}-m_{B}|<0.008$~GeV/$c^{2}$ for
both lepton modes and
$-0.055$~GeV$<\Delta E<0.035$~GeV ($|\Delta E|<0.035$~GeV) for the electron
(muon) mode.
We make distinct selections on ${\cal{R}}_{\rm cont}$ and ${\cal{R}}_{B\overline{B}}$
for each decay mode to optimize the sensitivity for the region with $q^2$ less than 6~GeV/$c^2$.

%
%

To determine the signal yield, we perform an unbinned maximum-likelihood fit to the $M_{\mathrm{bc}}$ distribution of events with $M_{\rm bc}> 5.2$~GeV/$c^{2}$ and $-0.055$~GeV$<\Delta E<0.035$~GeV ($|\Delta E|<0.035$~GeV) for the electron (muon) mode.
We consider the signal, three cross-feed and four background components. The cross-feed components are correctly and incorrectly flavor tagged $\Kstll$ events (``cfcf'' and ``ifcf'', respectively) and $b \to s \ell^+ \ell^-$ processes other than $\Kstll$ ($\Xsll$ events). The four background components are dilepton background, $\Kstlh$ background, $\Ksthh$ background and $\psi$ background, where $h$ refers to a pion or a kaon. The dominant components in the signal box are signal and dilepton background. For the muon mode, $\Kstlh$ background is sub-dominant. The ratio of $\Kstlh$ to dilepton background is about 0.24 and 0.12 for muon and electron modes, respectively.

The expected number of signal events  is 
calculated as a function of $M_{\mathrm{bc}}$ using a Gaussian  signal distribution plus  background functions.
The mean and the width of the signal Gaussian 
are determined using observed $J/\psi K^{(*)}$ events.
A MC study shows that the width has no dependence  on the dilepton invariant mass.
The shape and normalization of cross-feed events are parameterized by a Gaussian and an ARGUS function~\cite{ARGUS} based on MC samples.

The background from semileptonic decays is parameterized by 
an ARGUS function.
The shape of the backgrounds are determined from large $B\bar{B}$ and continuum MC samples in which
each event contains at least one oppositely charged lepton pair.
The shape parameter obtained from MC is consistent with that
taken from a data sample of $B \to K^{(*)}e^{\pm}\mu^{\mp}$ candidates.
Since the $\Mbc$ shape for $\Kstlh$ background is similar to that for the dilepton background, we use the same shape function. 
The residual background from $J/\psi$ and $\psi'$ mesons that cannot
be removed by the $\psi^{(')}$ veto windows is estimated from a large MC sample of
$J/\psi$ and $\psi'$ inclusive events and parameterized by an ARGUS
function and a Gaussian. 
The background contribution due to misidentification of hadrons 
as leptons is  parameterized
by another ARGUS function and a Gaussian.
The ARGUS function represents the combinatorial background while
the Gaussian represents the component that forms a peak
in the $M_{\rm bc}$ distribution. 
The shape and normalization of this background are 
fixed using the $B\to K^{(*)}h^+h^-$ data sample (where $h$ refers to a pion or kaon).
All $K^{(*)} h^{+} h^{-}$ combinations are weighted by the momentum-
and polar angle-dependent probability of misidentifying $K^{(*)} h^+h^-$ as $K^{(*)} \ell^+\ell^-$.
This study yields $3.8 \pm 0.4$ and $2.5 \pm 0.3$ peaking events in the signal box for $K h^+ h^-$ and $K^* h^+ h^-$ , respectively.
Other backgrounds with misidentified leptons are negligible.
The probability density function(PDF) for the $\Mbc$ fit is expressed as follows:
\begin{eqnarray}
P(\Mbc) &=& f_{\sig}G_{\sig}(\Mbc)\nonumber \\
        &+& f_{\sig}f_{\rm cfcf}\left (f^G_{\rm cfcf}G_{\rm cfcf}(\Mbc) + (1-f^G_{\rm cfcf})A_{\rm cfcf}(\Mbc)\right )\nonumber \\
        &+& f_{\sig}f_{\rm ifcf}\left (f^G_{\rm ifcf}G_{\rm ifcf}(\Mbc) + (1-f^G_{\rm ifcf})A_{\rm ifcf}(\Mbc)\right ) \nonumber \\
        &+& f_{\Xs2ll}\left (f^G_{\Xs2ll}G_{\Xs2ll}(\Mbc) + (1-f^G_{\Xs2ll})A_{\Xs2ll}(\Mbc)\right ) \nonumber \\
        &+& f_{\rm dilep}A_{\rm dilep}(\Mbc) \nonumber \\
        &+& f_{\Kstarhh}\left (f^G_{\Kstarhh}G_{\Kstarhh}(\Mbc) + (1-f^G_{\Kstarhh})A_{\Kstarhh}(\Mbc)\right )\nonumber \\
        &+& f_{\psi}\left (f^G_{\psi}G_{\psi}(\Mbc) + (1-f^G_{\psi})A_{\psi}(\Mbc)\right ),
\end{eqnarray}
where $f$ is the fraction of each of signal and backgrounds, $G$ is a Gaussian and $A$ is an ARGUS function. The $f_{\sig}$ and $f_{\rm dilep}$ are floated in the fit. 

\begin{figure}[htbp]
\begin{center}
  \includegraphics[scale=1.0]{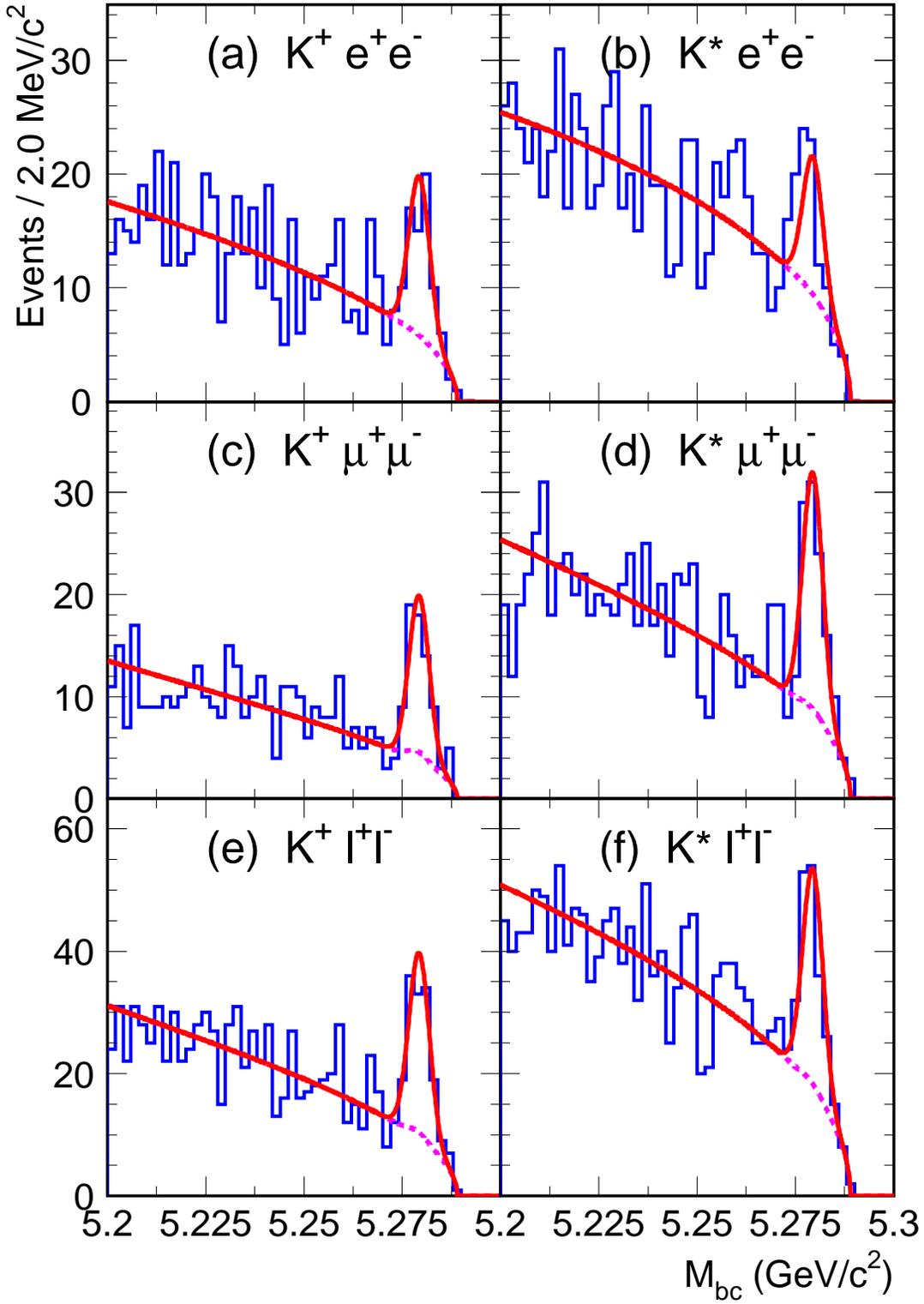}
\end{center}
\caption{$M_{\rm bc}$ distributions for (a) $B\to K^+ e^{+} e^{-}$, (b)
$B \to K^{*} e^{+} e^{-}$, (c) $B\to K^+ \mu^{+} \mu^{-}$, (d) $B \to
K^{*} \mu^{+} \mu^{-}$, (e) $B\to K^+ \ell^{+} \ell^{-}$ and (f) $B \to
K^{*} \ell^{+} \ell^{-}$ samples. The solid
and dashed curves are the fit results of the total and background
contributions.}
\label{fig:mbcfit}
\end{figure}

Figure~\ref{fig:mbcfit} shows the fit result. We obtain $96.0 \pm 12.0$ and $113.6 \pm 13.0$ signal events for $K^+\ell^+\ell^-$ and $\Kstll$, respectively. Table~\ref{tab:mbcfit} summarizes the signal yield for each subdecay mode.
\begin{table}[htbp]
 \begin{center}
 \begin{tabular}{lc} \hline \hline 
  {Mode}                       & Signal yield  \\ \hline
$K^+ e^+ e^-$                  & $46.6\pm8.7$  \\
$(K^+\pi^-)^{*0}e^+ e^-$       & $22.4\pm7.0$  \\
$(K^0_S\pi^+)^{*+}e^+ e^-$       & $10.9\pm4.4$  \\
$(K^+\pi^0)^{*+}e^+ e^-$       & $ 5.8\pm4.0$   \\\hline
$K^+\mu^+\mu^-$                & $49.4\pm8.2$  \\
$(K^+\pi^-)^{*0}\mu^+\mu^-$    & $56.6\pm8.8$   \\
$(K^0_S\pi^+)^{*+}\mu^+\mu^- $   & $ 7.7\pm4.0$   \\
$(K^+\pi^0)^{*+}\mu^+\mu^- $   & $10.2\pm4.1$   \\ \hline \hline
  \end{tabular}
   \end{center}
  \caption{%
    Signal yields obtained from the
    $M_{\mathrm{bc}}$ fit. The errors are  statistical only. }
\label{tab:mbcfit}
\end{table}

%
%
To extract Wilson coefficients, we perform an unbinned maximum likelihood fit to the normalized double differential decay width $(1/\Gamma) d\Gamma/dq^2d\cos\theta$ for $B \to \Kstll$~\cite{bib:abhh}, where $\cos\theta$ is the cosine of the angle between negative~(positive) charged lepton and $B^0$ or $B^+$~($\ol{B}^0$ or $B^-$) meson momenta in the dilepton rest frame. The forward-backward asymmetry is defined as
\begin{eqnarray}
A_{\rm FB}(q^2) = \frac{\Gamma(q^2,\cos\theta>0) - \Gamma(q^2,\cos\theta<0)}{\Gamma(q^2,\cos\theta>0) + \Gamma(q^2,\cos\theta<0)}.
\end{eqnarray} 
We fix all parameters in the $C_i^{\rm eff}$ to be SM values except the leading coefficients $A_7$, $A_9$ and $A_{10}$. In the SM at renormalization scale $\mu = 2.5$~GeV, the values of the leading coefficients are $A_7 = -0.330$, $A_9 =4.069$ and $A_{10} = -4.213$~\cite{ALGH,WC}.
In the double differential decay width, there are a few additional parameters to be fixed. These include the form factor parameters and the bottom quark mass $m_b$. We choose the form factor model of Ali {\it et al.}~\cite{ALGH,bib:abhh} and the bottom quark mass of $4.8$~GeV$/c^2$ for the nominal fit. The renormalization scale $\mu$ is set at 2.5~GeV as suggested by Ref.~\cite{ALGH}.

The PDF used for the fit consists of terms describing signal, cross-feeds and backgrounds:
\begin{eqnarray}
P(q^2,\cos\theta,\Mbc) &=& f_{\sig}(\Mbc)\epsilon_{\sig}^{\rm DS}(q^2,\cos\theta)\frac{d^2\Gamma}{dq^2 d\cos\theta}(q^2,\cos\theta;A_7,A_9,A_{10})/N_{\sig}^{\rm DS}(A_7,A_9,A_{10})\nonumber \\
		&+& f_{\rm cfcf}(\Mbc)\epsilon_{\rm cfcf}(q^2,\cos\theta)\frac{d^2\Gamma}{dq^2 d\cos\theta}(q^2,\cos\theta;A_7,A_9,A_{10})/N_{\rm cfcf}(A_7,A_9,A_{10})\nonumber \\
		&+& f_{\rm ifcf}(\Mbc)\epsilon_{\rm ifcf}(q^2,\cos\theta)\frac{d^2\Gamma}{dq^2 d\cos\theta}(q^2,-\cos\theta;A_7,A_9,A_{10})/N_{\rm ifcf}(A_7,A_9,A_{10})\nonumber  \\
		&+& f_{\Xs2ll}(\Mbc){\cal P}_{\Xs2ll}(q^2,\cos\theta)\nonumber \\
		&+& f_{\rm dilep}(\Mbc)\Big\{(1-f_{\Kstarlh}){\cal P}_{\rm dilep}(q^2,\cos\theta)+f_{\Kstarlh}{\cal P}_{\Kstarlh}(q^2,\cos\theta)\Big\} \nonumber \\
		&+& f_{\Kstarhh}(\Mbc){\cal P}_{\Kstarhh}(q^2,\cos\theta)\nonumber \\
		&+& f_{\psi}(\Mbc){\cal P}_{\psi}(q^2,\cos\theta),
\end{eqnarray}
where $f$ and $\cal{P}$ are the event by event fraction and probability density function for each components, respectively. $N$ is the normalization factor and $\epsilon$ is the efficinecy function. Each of the fractions ($f_{\rm sig}$ etc.) are determined mode by mode from the $\Mbc$ fits except $f_{\Kstarlh}$, the fraction of $\Kstlh$ background within the dilepton background component, which is determined from the MC samples. 
The efficiency functions $\epsilon$ and background distributions ${\cal P}$ for $\Xsll$ event, dilepton background, $\Kstlh$ background and $\psi$ background are obtained from large MC samples. 
Different efficiency functions are used for DS-I and DS-II since the acceptances are slightly different.
A common background PDF used for both data sets with the averaged parameters.
The Wilson coefficients in the PDF for $\Xsll$ are fixed at their SM values since this background's contribution is negligibly small. 
The $\Ksthh$ background shape ${\cal P}_{\Ksthh}$ is obtained from $\Ksthh$ data, applying momentum- and angular-dependent lepton ID fake rate corrections. 

The input parameters in the fit are $q^2$, $\cos\theta$ and $\Mbc$. Only $A_9/A_7$ and $A_{10}/A_7$ are allowed to float in the fit. We use smoothed histograms for efficiency and dilepton background distributions, and unsmoothed histograms for other event categories.
The background PDF shapes are found to agree with the the distributions in the data sideband well.

\begin{figure}[htbp]
\begin{center}
  \includegraphics[scale=0.8]{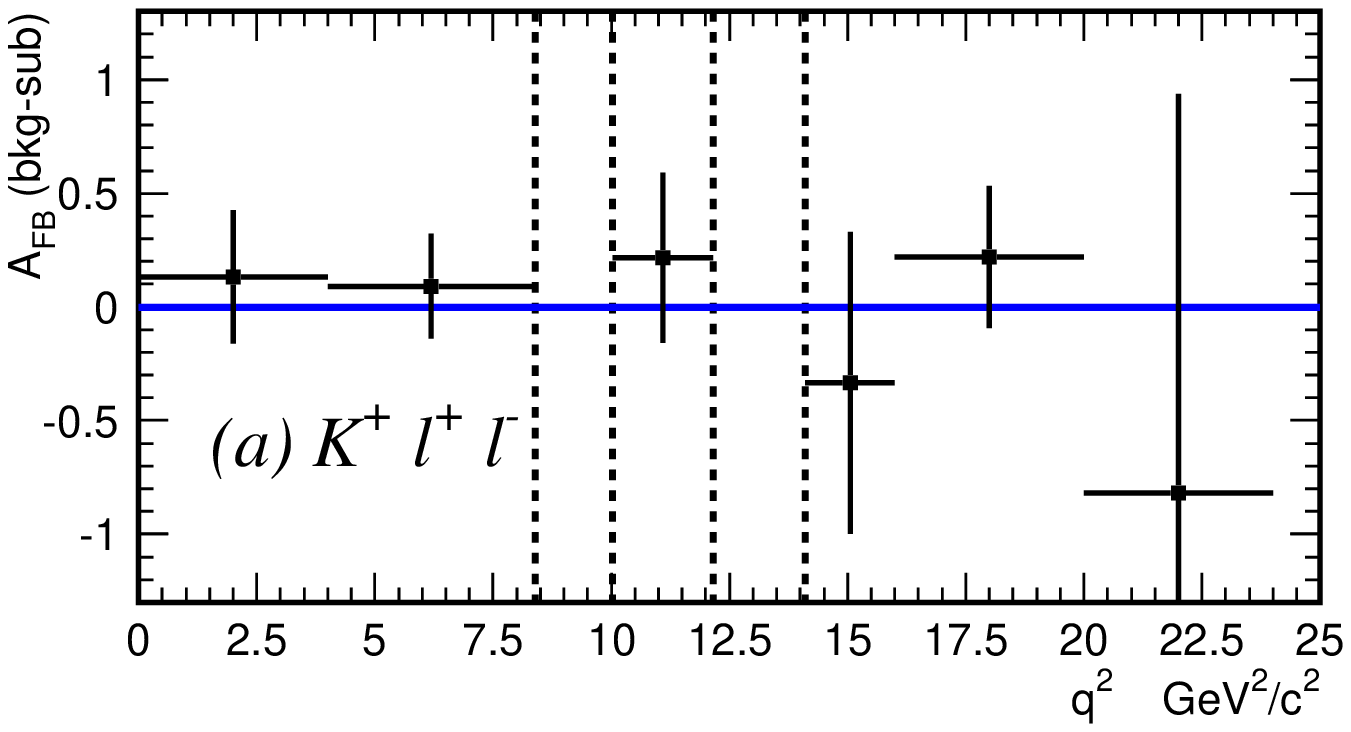}
  \includegraphics[scale=0.8]{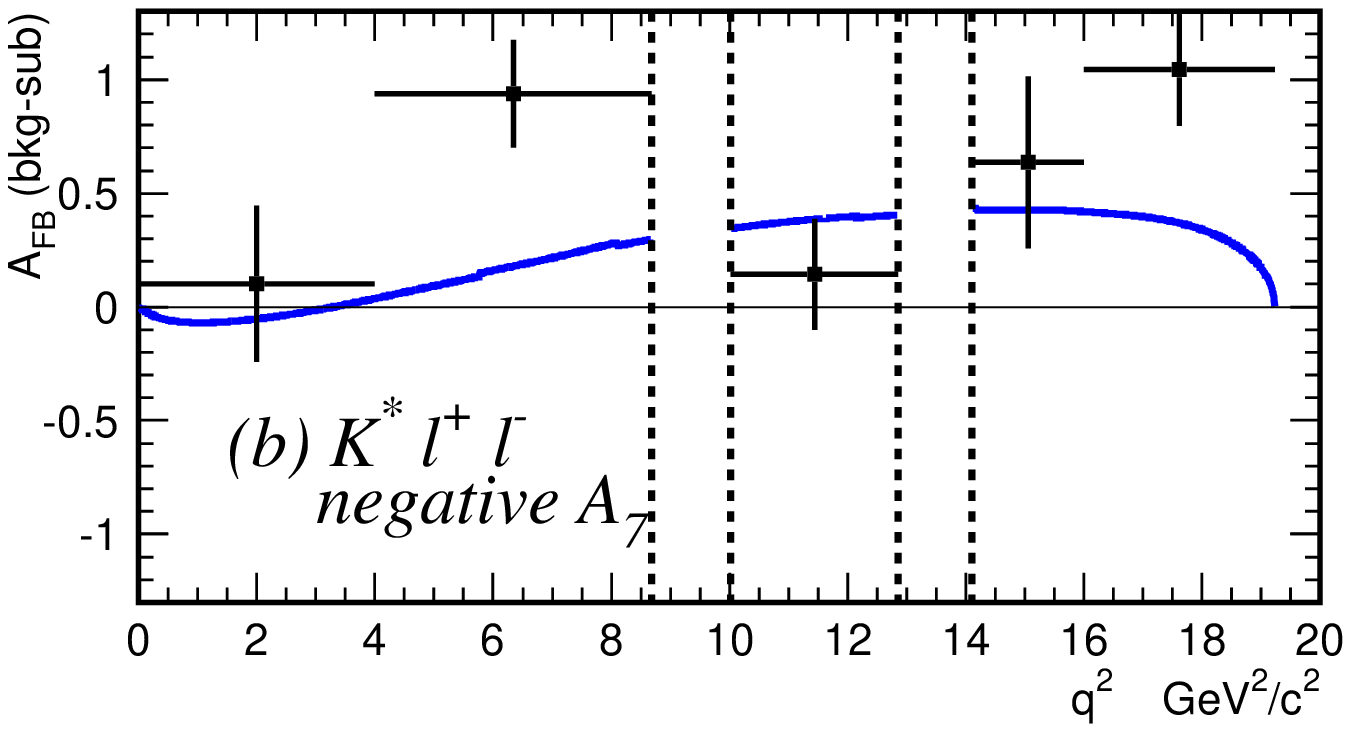}
  \includegraphics[scale=0.8]{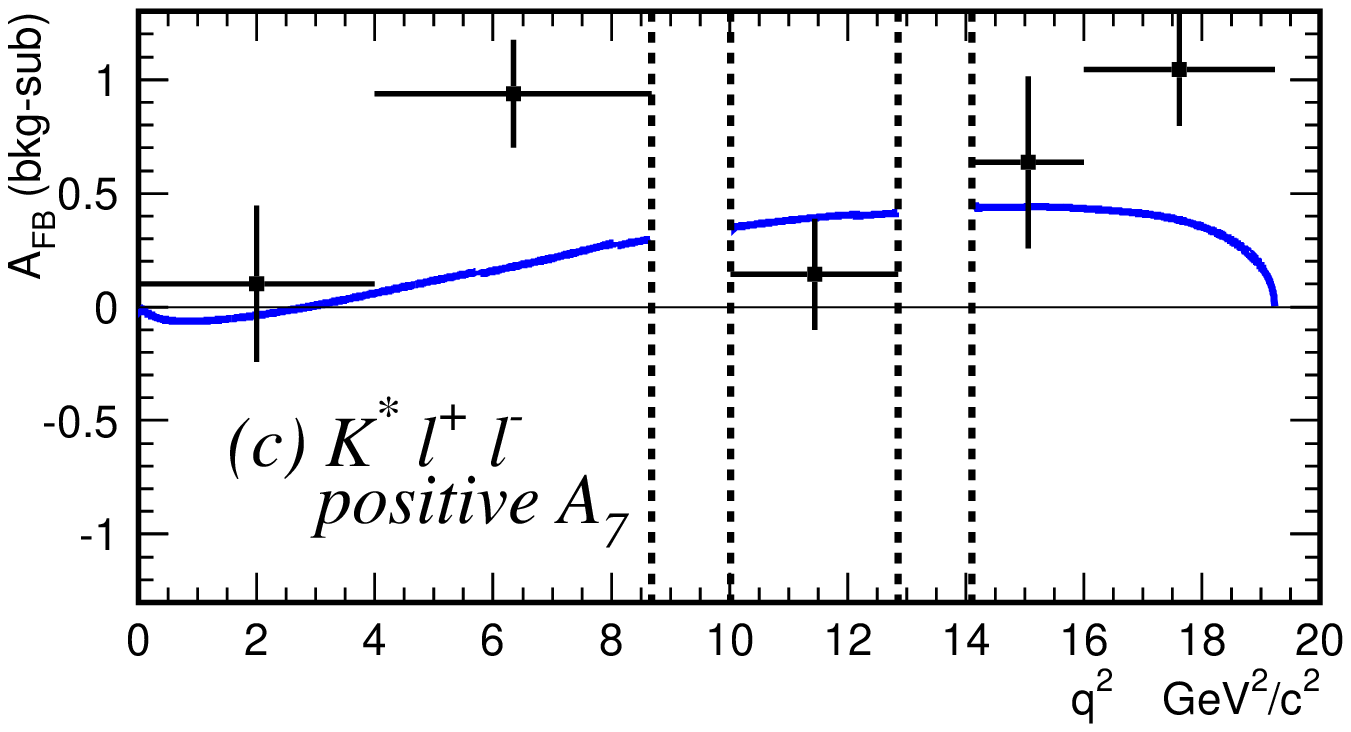}
\end{center}
\caption{Forward-backward asymmetry distributions for $\Kll$ and $\Kstll$. The points show the results obtained from $\Mbc$ fits while the curves show the fit results.}
\label{fig:afb}
\end{figure}

Prior to the fit, we measure the background subtracted integrated asymmetry. 
To check the possible bias due to the event selection, we perform the null asymmetry test with the $\Kll$ control sample.
We obtain 
\begin{eqnarray}
A_{\rm FB}^{\rm bkg-sub}(B \to K^+ \ell^+ \ell^-) &=& 0.09 \pm 0.14{\rm(stat.)}.\\
\end{eqnarray}
Note that $A_{\rm FB}^{\rm bkg-sub}$ is not corrected for the $q^2$-dependent efficiency.
The result for $B\to K^+ \ell^+ \ell^-$ is consistent with zero as expected (Fig.\ref{fig:afb}(a)). We thus confirm that the selection
criteria do not produce any bias. On the other hand, a large asymmetry is observed for $B \to \Kstll$ (Fig.\ref{fig:afb}(b),(c)):
\begin{eqnarray}
A_{\rm FB}^{\rm bkg-sub}(B \to \Kstll) &=& 0.56 \pm 0.13{\rm(stat.)}.
\end{eqnarray}

We fit the $q^2$ and $\cos\theta$ distribution for $\Kstll$ with the PDF described in the previous section.
Using $A_7 = -0.330$, the best fit Wilson coefficient ratios are
\begin{eqnarray}
{A_9}/{A_7}    &=& -15.3 \PM{3.4}{4.8},\nonumber \\
{A_{10}}/{A_7} &=& 10.3 \PM{5.2}{3.5}\,;
\label{eq:nega}
\end{eqnarray}
while, with $A_7 = +0.330$, the best fit ratios are
\begin{eqnarray}
{A_9}/{A_7}    &=& -16.3 \PM{3.7}{5.7},\nonumber \\
{A_{10}}/{A_7} &=&  11.1 \PM{6.0}{3.9}.
\label{eq:posi}
\end{eqnarray}
Figure~\ref{fig:afb} shows the fit results projected on the background-subtracted forward-backward asymmetry distribution.

We study the systematic error due to uncertainty in the physics parameters, finite $q^2$ resolution, 
efficiencies for low momentum particles and the signal fraction.
We change the $A_7$ value within the range allowed by the branching fraction of $B \to X_s \gamma$~\cite{bib:hiller} to obtain the corresponding systematic error.
The bottom quark mass $m_b$ is varied by $\pm0.2$~GeV. The systematic uncertainty associated with the choise of form factor model is taken from the difference in fit results using the models of Ali {\it et al.} and Melikhov {\it et al.}~\cite{bib:mns}.
We ignore smearing in $q^2$ in the fits, although the actual resolution is about 0.7\%; the effect is measured using a toy MC study.
The efficiencies to find and identify low momentum particles are slightly better in MC events, so
we simultaneously reduce the efficiency for pion with $p<0.3$~GeV, electron having $p<0.7$~GeV and muon with $p<1.0$~GeV to obtain revised efficinecy functions for signal and background PDFs; we then refit the data with the new PDFs. The change in the fit results correspond to the systematic error associated with the identification efficinecy.
The signal fraction is determined from a fit to $\Mbc$ distribution. We change the PDF shape parameters and take the difference as a systematic error: the parameters are modified by $\pm1\sigma$ for signal and dilepton background, by 100\% for cross-feed events and $\psi$ background, and 10\%(40\%) for $\Ksthh$ background for muon(electron) modes. For the uncertainty in $\Kstlh$ background, we change the fraction of $\Kstlh$ to dilepton backround, $f_{\Kstlh}$, by 20\%.
Table~\ref{tab:syst} summarizes the contributions to the systematic error. The dominant contribution is from model dependence. 
\begin{table}[htbp]
 \begin{center}
 \begin{tabular}{lcccccc} \hline \hline
source&
\multicolumn{2}{c}{negative $A_7$ solution}&
\multicolumn{2}{c}{positive $A_7$ solution} &  \\ 
                    & $A_9/A_7$         & $A_{10}/A_7$      & $A_9/A_7$         & $A_{10}/A_7$ \\ \hline
$A_7$         	    & $\PM{0.29}{0.03}$ & $\PM{0.01}{0.03}$ & $\PM{0.13}{0.27}$ & $\PM{0.36}{0.15}$ \\
$m_b$               & $\PM{0.69}{0.68}$ & $\PM{0.45}{0.46}$ & $\pm 0.63$        & $\pm0.42$ \\
model dependence    & $\pm0.66$         & $\pm1.72$         & $\pm1.04$         & $\pm2.23$ \\
$q^2$ resolution    & $\pm0.28$         & $\pm0.39$         & $\pm0.28$         & $\pm0.39$\\
efficiency          & $\pm0.08$         & $\pm0.03$         & $\pm0.10$         & $\pm0.06$\\
signal fraction     & $\PM{0.43}{0.47}$ & $\PM{0.22}{0.33}$ & $\PM{0.43}{0.46}$ & $\PM{0.37}{0.40}$\\ \hline
total               & $\PM{1.12}{1.10}$ & $\PM{1.83}{1.84}$ & $\PM{1.33}{1.36}$ & $\PM{2.36}{2.34}$ \\ \hline \hline
  \end{tabular}
  \end{center}
  \caption{%
    Systematic error for each source. }
\label{tab:syst}
\end{table}

\begin{figure}[htbp]
\begin{center}
  \includegraphics[scale=0.75]{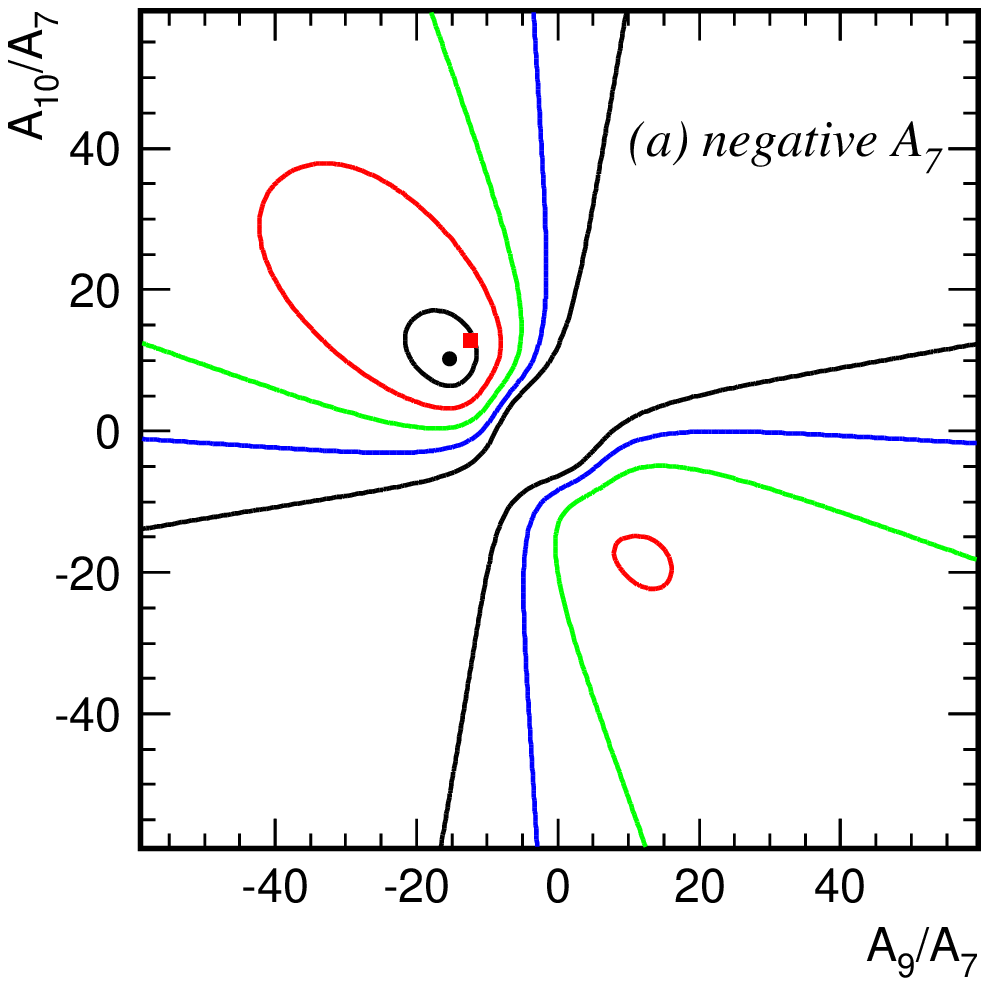}
  \includegraphics[scale=0.75]{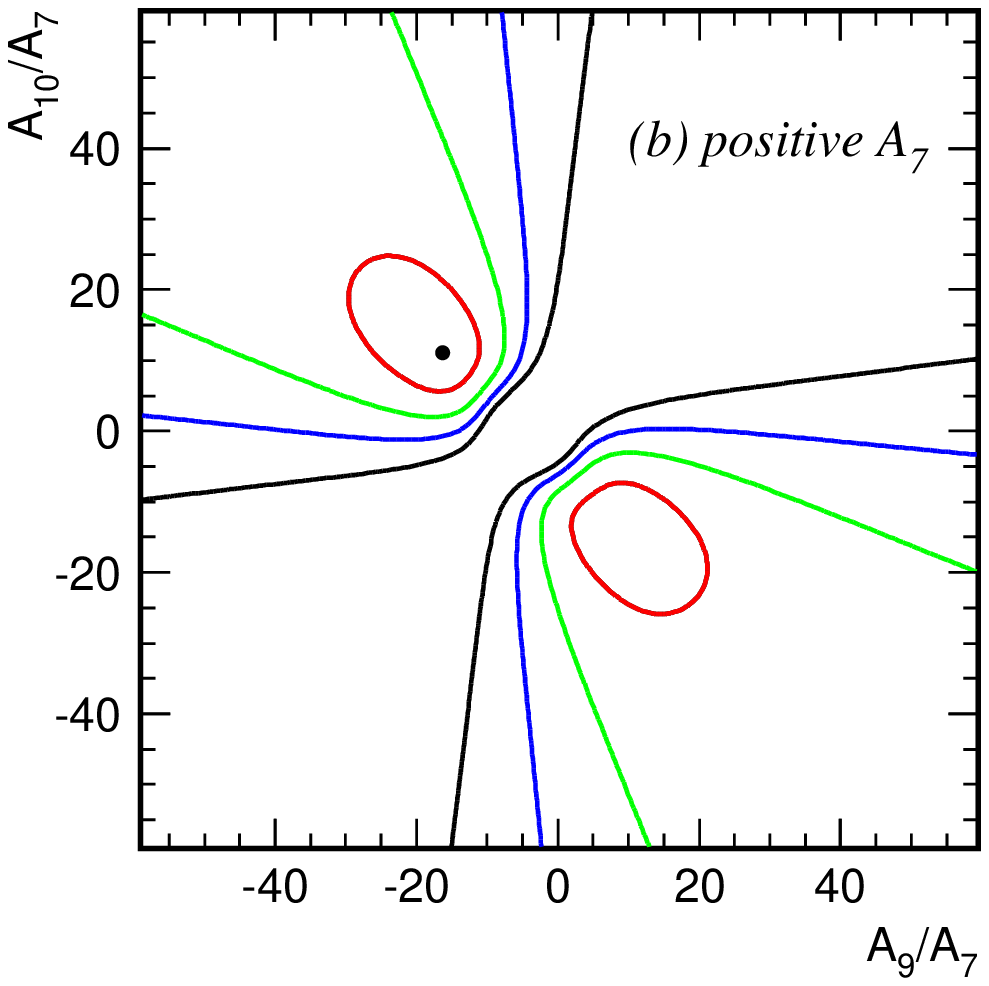}
\end{center}
\caption{Confidence level contours for negative and positive $A_7$. Black, red, green, blue and black curves show 1 to 5 $\sigma$ contours. Black and red points show the best fit and the SM value.}
\label{fig:cl}
\end{figure}

\begin{figure}[htbp]
\begin{center}
  \includegraphics[scale=1.0]{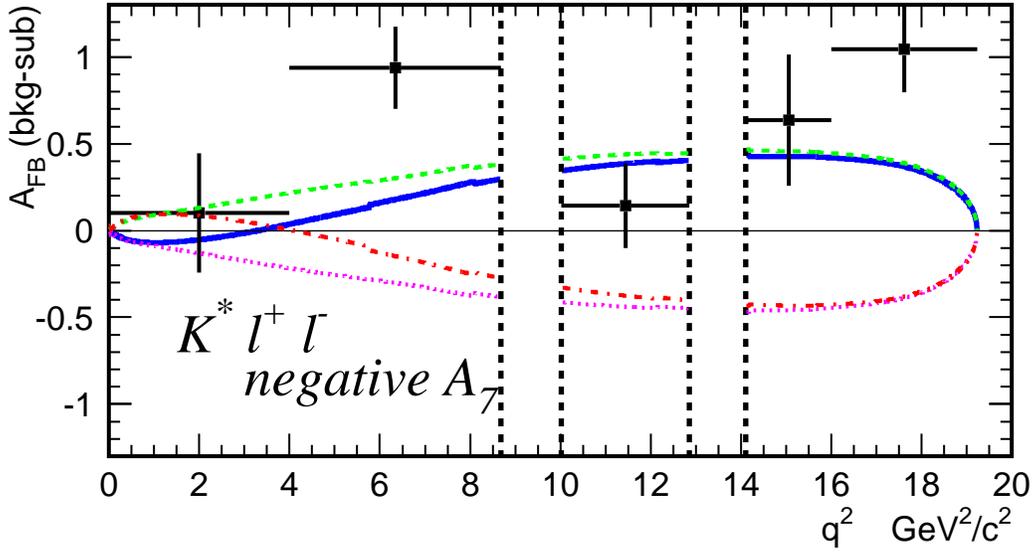}
\end{center}
\caption{Fit results for the negative $A_7$ solution projected to forward-backward asymmetry(solid blue) and forward-backward asymmetry curves with several input parameter including efficiency effect; $A_7A_{10}$ sign flipped (dashed green), both $A_7A_{10}$ and $A_9A_{10}$ signs flipped (dash-dot red) and $A_9A_{10}$ sign flipped(dotted magenta) to SM value. The new physics scenarios shown by the red and magenta curves are excluded.}
\label{fig:afbex}
\end{figure}

%
%
\newpage
The best-fit results for the Wilson coefficient ratios for negative $A_7$,
\begin{eqnarray}
{A_9}/{A_7}    &=& -15.3 \PM{3.4}{4.8} \pm 1.1,\nonumber \\
{A_{10}}/{A_7} &=& 10.3 \PM{5.2}{3.5} \pm 1.8\,;
\label{eq:negasyst}
\end{eqnarray}
and positive $A_7$
\begin{eqnarray}
{A_9}/{A_7}    &=& -16.3 \PM{3.7}{5.7} \pm 1.4,\nonumber \\
{A_{10}}/{A_7} &=&  11.1 \PM{6.0}{3.9} \pm 2.4,
\label{eq:posisyst}
\end{eqnarray}
are consistent with the SM values ${A_9}/{A_7}=-12.3$ and ${A_{10}}/{A_7}=12.8$.
In Fig.~\ref{fig:cl}, we draw contour lines in the $A_9/A_7$-$A_{10}/A_7$ plane based on the fit likelihood smeared by systematic error (assuming Gaussian).
We also calculate an interval on $A_9A_{10}/A_7^2$ at 95\% CL for any $A_7$ value,
\begin{eqnarray}
-1401 < {A_9}{A_{10}}/{A_7^2} < -26.4.
\end{eqnarray}
We determine the sign of ${A_9}{A_{10}}$ to be negative, and exclude solutions in the first or third quadrant with more than 95\% CL.
Both second and fourth quadrant solutions are allowed, so we cannot determine the sign of $A_7A_{10}$ yet.
Figure~\ref{fig:afbex} shows the comparison between the fit results for the negative $A_7$ solution projected on the forward-backward asymmetry and the forward-backward asymmetry distributions for several input parameters. We exclude new physics scenarios shown by the red and magenta curves, which have positive ${A_9}{A_{10}}$.

%
%
In summary, we have measured for the first time the ratios of Wilson coefficients in $B \to \Kstll$ by fitting to a normalized double differential decay width. We observe a large forward-backward asymmetry in $B \to \Kstll$.
The fit result is consistent with the SM prediction and also with the case where the sign of $A_7A_{10}$ flipped. We exclude positive $A_9A_{10}$ solutions at more than 95\% CL.

%
%

We would like to thank Gudrun Hiller for her invaluable suggestions.
We thank the KEKB group for the excellent operation of the
accelerator, the KEK cryogenics group for the efficient
operation of the solenoid, and the KEK computer group and
the National Institute of Informatics for valuable computing
and Super-SINET network support. We acknowledge support from
the Ministry of Education, Culture, Sports, Science, and
Technology of Japan and the Japan Society for the Promotion
of Science; the Australian Research Council and the
Australian Department of Education, Science and Training;
the National Science Foundation of China under contract
No.~10175071; the Department of Science and Technology of
India; the BK21 program of the Ministry of Education of
Korea and the CHEP SRC program of the Korea Science and
Engineering Foundation; the Polish State Committee for
Scientific Research under contract No.~2P03B 01324; the
Ministry of Science and Technology of the Russian
Federation; the Ministry of Higher Education, 
Science and Technology of the Republic of Slovenia;  
the Swiss National Science Foundation; the National Science Council and
the Ministry of Education of Taiwan; and the U.S.\
Department of Energy.


%


\begin{thebibliography}{99}
\bibitem{aster}
Throughout this paper, variables denoted with an asterisk are
calculated in the $\Upsilon(4S)$ rest frame.
\end{thebibliography}
\end{document}